\documentclass[letterpaper, 10 pt, conference]{ieeeconf}  

\IEEEoverridecommandlockouts                              
\usepackage{amsmath} 
\usepackage{amssymb}  

\usepackage{url}
\usepackage[ruled, vlined, linesnumbered]{algorithm2e}
\usepackage{verbatim} 
\usepackage{soul, color}
\usepackage{lmodern}
\usepackage{fancyhdr}
\usepackage[utf8]{inputenc}
\usepackage{fourier} 
\usepackage{array}
\usepackage{makecell}
\usepackage{scicite}
\usepackage{float}
\usepackage{upgreek}
\usepackage{graphicx}
\usepackage{subfigure}

\UseRawInputEncoding

\SetNlSty{large}{}{:}

\makeatletter

\newcommand{\Rom}[1]{\expandafter\@slowromancap\romannumeral #1@}
\makeatother

\title{\LARGE
Realization of inverse-design magnonic logic gates
}

\author{Noura Zenbaa,$^{1,2\ast}$ Fabian Majcen,$^{1,2}$ Claas Abert,$^{1,3}$ Florian~Bruckner,$^{1,3}$ \\ Norbert~J.~Mauser,$^{3,4}$ Thomas Schrefl,$^{3,5}$ Qi Wang,$^{6}$ Dieter Suess,$^{1,3}$ Andrii V. Chumak,$^{1,3\ast}$\\
\\
\normalsize{$^{1}$University of Vienna, Faculty of Physics, Vienna 1090, Austria}\\
\normalsize{$^{2}$University of Vienna, Vienna Doctoral School in Physics, Vienna 1090, Austria}\\
\normalsize{$^{3}$Research Platform MMM "Mathematics-Magnetism-Materials", University of Vienna, Vienna 1090, Austria.}\\
\normalsize{$^{4}$Faculty of Mathematics, University of Vienna, Vienna 1090, Austria.}\\
\normalsize{$^{5}$Center for Modelling and Simulation, Donau-Universit\"at Krems, Wiener Neustadt, 2700, Austria.}\\
\normalsize{$^{6}$School of Physics, Hubei Key Laboratory of Gravitation and Quantum Physics,}\\ \normalsize{Institute for Quantum Science and Engineering, Huazhong University of Science and Technology, Wuhan, 430074, China.}
\\
\normalsize{$^\ast$To whom correspondence should be addressed; E-mail: noura.zenbaa@univie.ac.at, andrii.chumak@univie.ac.at}}

\date{}

\begin{document}

\maketitle

\thispagestyle{plain}
\pagestyle{plain}

\begin{abstract}

Magnonic logic gates represent a crucial step toward realizing fully magnonic data processing systems without reliance on conventional electronic or photonic elements. Recently, a universal and reconfigurable inverse-design device has been developed, featuring a 7$\times$7 array of independent current loops that create local inhomogeneous magnetic fields to scatter spin waves in a Yttrium-Iron-Garnet film. While initially used for linear RF components, we now demonstrate key non-linear logic gates, NOT, OR, NOR, AND, NAND, and a half-adder, sufficient for building a full processor. In this system, binary data ("0" and "1") are encoded in the spin-wave amplitude. The contrast ratio, representing the difference between logic states, achieved values of 34, 53.9, 11.8, 19.7, 17, and 9.8 dB for these gates, respectively.\\

\end{abstract}


\section{INTRODUCTION}\label{sec1}

Logic gates are essential components in modern digital electronics, enabling complex computational tasks and serving as the backbone of modern semiconductor information processing systems. The growing demand for faster and more efficient computing has driven the development of increasingly sophisticated logic gates, which are critical for advancing computer architectures and integrated circuits. However, as traditional electronic logic gates approach their physical and thermal limits, there is increasing interest in exploring alternative technologies that can offer improved performance and efficiency. Magnonic logic gates, which use spin waves instead of charge carriers, present a promising alternative. Magnonics, the field focused on utilizing magnons, the quanta of spin waves, for data processing and transmission, excels at promoting low-power wave computing~\cite{barman_2021_2021, iwaba_spin-wave_2021, chumak_advances_2022,khivintsev_spin_2022}. Spin waves operate across a wide frequency range, from sub-GHz to the THz regime~\cite{kampfrath_coherent_2011, wu_high-performance_2017}, and exhibit intrinsic nonlinear properties that can be harnessed for logic functions~\cite{krivosik_hamiltonian_2010, chumak_magnon_2014, sadovnikov_nonlinear_2016}. For these reasons, magnetic logic gates can potentially overcome the limitations faced by conventional electronics and enable next-generation data processing technologies~\cite{khitun_magnonic_2010, talmelli_reconfigurable_2020,adelmann_spintronic_2023, hayashi_enhanced_2024}.

Although various approaches have successfully demonstrated magnonic logic gates, including those based on a Mach-Zehnder interferometer to realize a NOT gate~\cite{kostylev_spin-wave_2005}, XNOR and NAND~\cite{schneider_realization_2008}, and XOR~\cite{ustinov_nonlinear_2019}, other approaches use dipolar coupling between magnetic waveguides to realize magnetic logic elements~\cite{hayashi_enhanced_2024}, like the numerically demonstrated half-adder based on two directional couplers~\cite{wang_magnonic_2020}. Spin-wave interference~\cite{goto_three_2019, schulz_realization_2023} and magnonic crystals~\cite{nikitin_spin-wave_2015} have also been employed to demonstrate XNOR, magnonic analog adders and AND gates, respectively. Moreover, recent work on complementary magnon transistors~\cite{chen_complementary_2023} has shown promise for the development of spin-wave
processors, while the development of spin-wave-based Ising machines~\cite{litvinenko_spinwave_2023} leverage spin-wave interference for solving combinatorial optimization problems. Innovations such as the SW-based 4:2 compressor utilizing three- and five-input majority gates~\cite{mahmoud_spin_2022} further expand the computational capabilities of spin-wave logic. Finally, the concept of a magnonic co-processor~\cite{balynsky_magnonic_2023} is paving the road towards the novel type of combinatorial logic devices. Despite these advances, most designs involve time-consuming steps, often including micromagnetic simulations, followed by precise fabrication and characterization~\cite{mahmoud_introduction_2020, garlando_numerical_2023}. Moreover, they are typically restricted to specific frequency ranges and perform only one functionality, limiting their flexibility.

A newly adapted concept in the field of magnonics, known as inverse design, can be leveraged to significantly shorten the design process by automating the optimization of structures, thus reducing both time and effort while ensuring higher precision in achieving desired functionalities. This concept relies on an objective-first approach that defines a functionality and combines it with a feedback-loop optimization process. A few inverse-design magnonic functionalities have been demonstrated both numerically~\cite{wang_inverse-design_2021, papp_nanoscale_2021} and experimentally~\cite{kiechle_experimental_2022}. All-optical inverse-design logic gates were recently shown only numerically in the field of photonics~\cite{neseli_inverse_2022, wang_ultra-broadband_2024, lan_inverse_2024}.

Our recent work demonstrated a reconfigurable universal inverse-design device that was used to realize multiple RF components, including a reconfigurable RF filter and a frequency de-multiplexer~\cite{zenbaa_experimental_2024}. However, all the presented results in this previous work were focused on linear functionalities, which do not apply to the implementation of binary and unconventional (neuromorphic or reservoir) computing.  

Here, we present the first experimental realization of inverse-design magnonic logic gates using the reconfigurable device described in~\cite{zenbaa_experimental_2024}. This universal device uses Forward-Volume Magnetostatic Spin-Wave (FVMSW) interference to achieve constructive or destructive outputs. The design uses a 7$\times$7 array of omega-shaped current loops, each generating an Oersted field of up to $\pm$3.5\,mT, to control the amplitude and phase of the spin wave propagating through inhomogeneous field regions. This creates complex interference patterns of nonlinear spin waves capable of solving versatile inverse design problems. A feedback loop optimization iterates over the 49 current values applied to the loops, using the spin-wave signal as feedback to achieve the desired functionality. We demonstrated the device's performance and potential by successfully implementing six key conceptual logic functions - NOT, OR, NOR, AND, NAND and half-adder - on the same platform. Logic states "0" and "1" are encoded in spin-wave amplitudes: signals above 90\% of the maximum amplitude represent "1", while those below 10\% represent "0". The device achieves high contrast ratios, up to 53.9\,dB for the OR gate, demonstrating its versatility and effectiveness.

\section{Results}\label{sec2}
\subsection{Universal inverse-design binary device}\label{subsec2}
\begin{figure*}[tbp!]
    \centering
    \includegraphics[width=\linewidth]{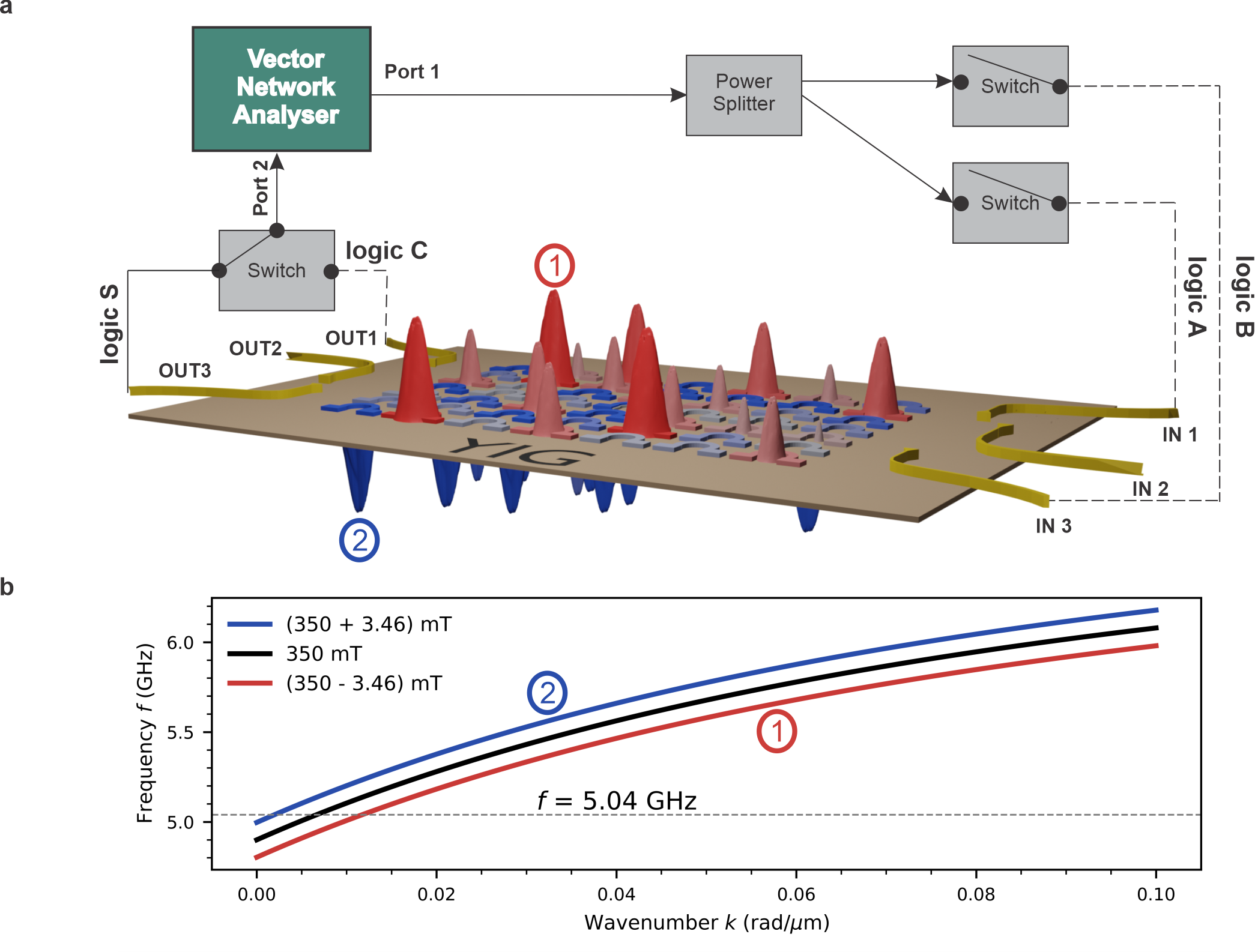}
    \caption{\textbf{The experimental setup of the half-adder logic functionality.} \textbf{a} The reconfigurable inverse-design device depicting the 7$\times$7 omega-shaped loop array placed on top a YIG film. It depicts the effect of the current applied to the loops as a field landscape generated locally on the film in red (film top) and blue (film bottom) for both current polarities, parallel or antiparallel to the applied out-pf-plane magnetic field. It includes the setup used to realise the most complicated logic gate presented in this work, a half-adder logic functionality, which involves a Vector Network Analyser (VNA) in a combination of a power splitter and two switches on the input side where all logic states can be achieved and one switch on the output side to measure transmission signals from both outputs. \textbf{b} Spin-wave dispersion relation at three different fields, 350\,mT which is the external bias field fixed throughout the measurements, (350 - 3.46)\,mT which is the field sensed by local spin waves at point \textbf{1} (indicated in panel a) when the Oersted field generated by a loop carrying a current of 400\,mA is antiparallel to the bias field and (350 + 3.46)\,mT at point \textbf{2} (indicated in panel a) when the field is generated by a loop carrying a current of the opposite polarity -400\,mA and therefore parallel to the bias field. It is important to note that the maximum generated field by an omega loop does not occur at the center of the loop but around the metallic loop.}
    \label{fig1}
\end{figure*}

The universal inverse-design magnonic device described in~\cite{zenbaa_experimental_2024} is depicted in Fig.~\ref{fig1}(a). The device is based on a YIG film and a design region of 49 omega-shaped current loops placed on top of it to manipulate the FVMSW propagation. The current loops are controlled independently and each can carry currents of $\pm$1\,A in 2048 steps using specially designed multi-channel current sources. The current was limited to only $\pm$400\,mA in 100\,mA steps in order to limit the device heating, which gives a total of 9 discrete current values. The physical effect of current-carrying loops is generating Oersted fields parallel or antiparallel to the applied out-of-plane external field as shown in Fig.~\ref{fig1}(a), which create local inhomogeneous magnetic field regions that force spin waves to change wavelength when propagating through them by shifting the spin-wave dispersion relations to higher or lower frequencies compared to the one at the applied bias field of 350\,mT. This effect is shown in Fig.~\ref{fig1}(b) for the maximum field inhomogeneity which is $\pm$3.46\,mT at the applied current of $\pm$400\,mA. The total degrees of freedom introduced by the design region in this case are $9^{49}\approx10^{47}$, sufficient to perform complex data processing. The device employs three input and three output transducers to cover a wide range of functionalities.

All logic gates presented in this work are optimised at a fixed single frequency point of 5.04\,GHz. The frequency was selected on the basis of having the same spin-wave transmission amplitude for all input-output combinations at 350\,mT - see Fig.~\ref{figS1}. The measurement process starts by sending a microwave signal to the input(s), using a power splitter and mechanical switches in case of multiple inputs. The input power is divided equally to keep the power of logic "1" the same, 25\,dBm. Using the mechanical switch, the input is put to open "1" or closed (no applied microwave signal) "0" to satisfy all combinations of the given truth table. In the case of a half-adder functionality, the setup used is shown in Fig.~\ref{fig1}(a). It requires two inputs A \& B and two outputs C \& S. For the inputs,it uses one power divider and two mechanical switches, one of which is connected to input A and the other to input B. For the state of "01" A is closed "0" while B is open "1" and vice versa for state "10". Both A and B are set to open "1" for the state "11" and both are closed for the state "00" (no input microwave signal sent to either). It applies the correct inputs to the corresponding state and runs the Direct Search (DS) optimisation process explained in~\cite{zenbaa_experimental_2024}. Then it measures two spin-wave amplitudes corresponding to outputs C \& S for each input logic combination at each current set configured by the optimisation algorithm via the use of another mechanical switch connected to the output transducers to switch between outputs C \& S. The optimisation algorithm iterates until it reaches a state that satisfies the pre-defined conditions. The conditions are defined as follows; in each iteration of the specific case of the half-adder, six transmission amplitudes in dB of the 5.04\,GHz spin-wave frequency are recorded. These amplitudes correspond to the transmissions at output C at input states "01", "10", and "11", and the other three correspond to output S at the same input states. These six transmission amplitudes are used to define the objective, where the maximum transmission of the six is considered a 100\% transmission and corresponds to the output logic "1". All other transmissions are compared to the maximum transmission value and translated into transmission ratios. The condition was unified for all logic gates presented where any output below 10\% of the maximum transmission is considered to be logic "0" and all outputs above 90\% of the maximum transmission are to be considered logic "1".  It is important to note that all the logic gate functionalities presented are nonlinear functionalities and therefore were realised in the spin-wave nonlinear regime at microwave power of 25\,dBm~\cite{wang_deeply_2023}.

\subsection{NOT gate}\label{subsec4}
\begin{figure*}[btp!]
    \centering
    \includegraphics[width=\linewidth]{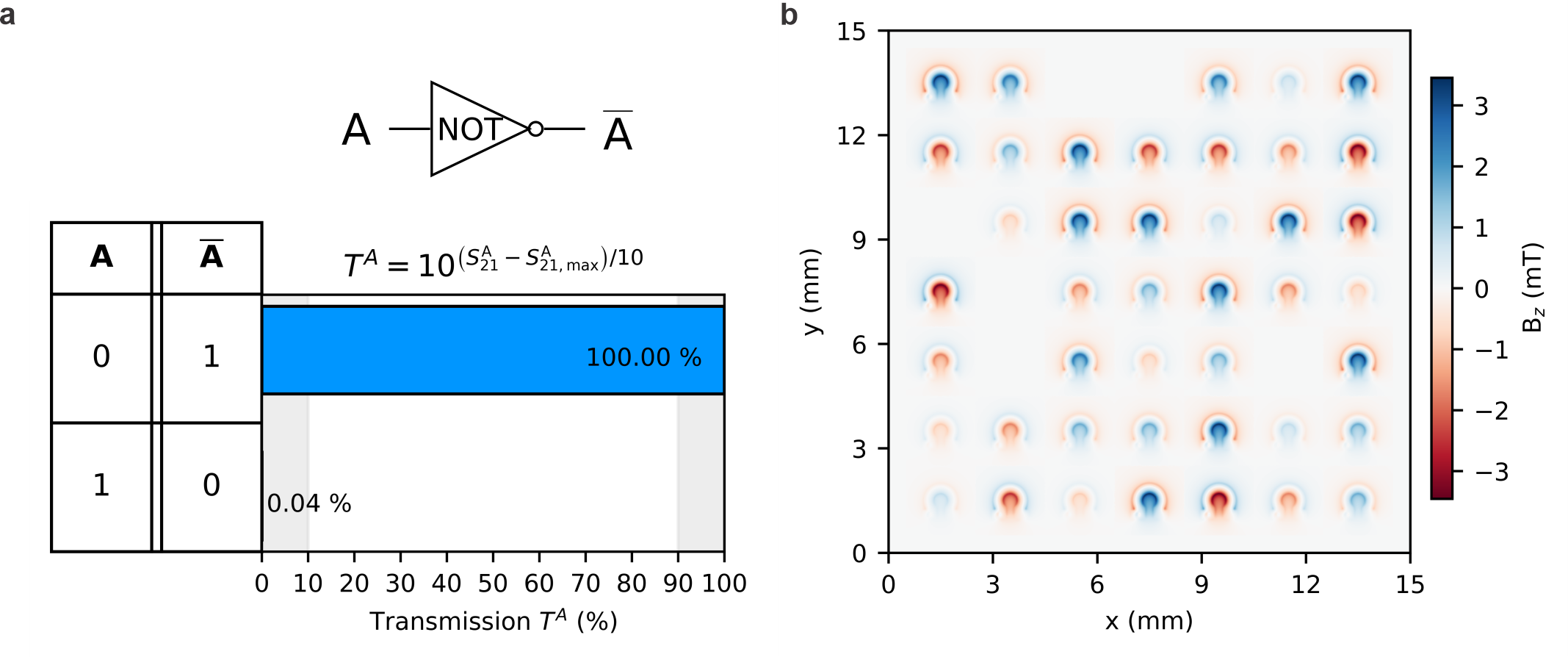}
    \caption{\textbf{The NOT logic functionality}. \textbf{a} The spin-wave transmission percentage of output $\overline{\mathrm{A}}$ at both logic states of input A. \textbf{c} A 2D color map representing the Oersted field generated by the omega-shaped loops to achieve results shown in panel a.}
    \label{fig2}
\end{figure*}

As a first example, we discuss the results of the NOT gate shown in Fig.~\ref{fig2}. The NOT gate, also known as an inverter, is fundamental in logic circuits and is one of the simplest logic gates. It has only one input and one output, and its function is simple: the output is always the opposite of the input. If the input is logic "1", the output will be logic "0", and vice versa. Input A was assigned to the IN1 transducer and output $\overline{\mathrm{A}}$ was assigned to OUT2 - see Fig.~\ref{fig1}(a). In order for the gate to output a logic "1" when the input logic is "0" (no input signal), there are two possible routes. First is encoding logic states in spin-wave phase instead of spin-wave amplitudes~\cite{khitun_multi-frequency_2012}. And second is by using a continuous feed-line F, which is the route taken in this case. The feed-line F is connected to IN3 and is always open while input A is set to both closed "0" and open "1" during the measurement procedure. Figure~\ref{fig2}(a) shows the optimised results that represent a NOT logic function using the objective function:
\begin{equation}
    O^{\mathrm{NOT}} = \left[T^{"0"} + (100 - T^{"1"})\right]/2,
\end{equation}

where $T^{"0"}$ \& $T^{"1"}$ are the transmission percentages at input A logic "0" and "1", respectively, with respect to the maximum transmission amplitude recorded per iteration. The objective function is commanded to be maximised, meaning that the closer it is to 100, the closer the optimiser is to converge and reach the functionality. The transmission percentage is calculated as follows:
\begin{equation}
    T^{A} = 10^{\frac{S^{\mathrm{A}}_{21} - S^{\mathrm{A}}_{\mathrm{21,max}}}{10}}\times 100
    \label{eqn2}
\end{equation}

This calculates linear power ratios between $S^{\mathrm{A}}_{21}$(dB) and $S^{\mathrm{A}}_{\mathrm{21,max}}$(dB), where $S^{\mathrm{A}}_{21}$ is the spin-wave transmission in dB from VNA port 1 to port 2 at a given current configuration where the input A can be closed "0" or open "1" and $S^{\mathrm{A}}_{\mathrm{21,max}}$ is the maximum transmission amplitude in dB of this iteration that occurs at either "0" or "1" input state. This means that this $S^{\mathrm{A}}_{\mathrm{21,max}}$ can vary throughout the iterations. Figure~\ref{fig2}(a) shows the percentage of output transmission for each given input state shown in the NOT truth table included on the left-hand side of the graph. It shows that at the input of the "0" logic it yields a transmission of 100\% that translates to the output logic "1" and only a transmission of 0.04\% at the input logic "1" that translates to the output logic "0". In this specific case 100\% transmission in Fig.~\ref{fig2}(a) corresponds to about -40\,dB, which is the maximum spin-wave transmission between the input and output transducers at 5.04\,GHz frequency at a specific current configuration applied, given the sample thickness, efficiency of the transducers, distance between transducers, microwave power applied, and microwave components used between input and output. The 0.04\% shown in Fig.~\ref{fig2}(a) corresponds to -74\,dB which is more than 1000 times less power than the logic "1" state in this case. This results in a contrast ratio of 34\,dB between the logic states "0" and "1". The contrast ratio is defined as $S^{"1"}_{21}\mathrm{(dB)} - S^{"0"}_{21}\mathrm{(dB)}$, where $S^{"1"}_{21}$ represents the minimum transmission output in dB at the output logic state "1", and $S^{"0"}_{21}$ represents the maximum transmission output in dB at the output logic state "0". The NOT gate functionality was achieved after 295 iterations. Considering the continuous feed-line approach used in this case, this gate also does half of the functionality of an XOR gate.

The Oersted field generated by the current configuration that satisfied the NOT gate functionality is shown in a color map in Fig.~\ref{fig2}(b). Since the maximum current was limited to $\pm$ 400\,mA, the maximum field generated was $\pm$3.46\,mT. The current conditions are applied for all presented data.

\subsection{OR and NOR gates}
\begin{figure*}[btp!]
    \centering
    \includegraphics[width=\linewidth]{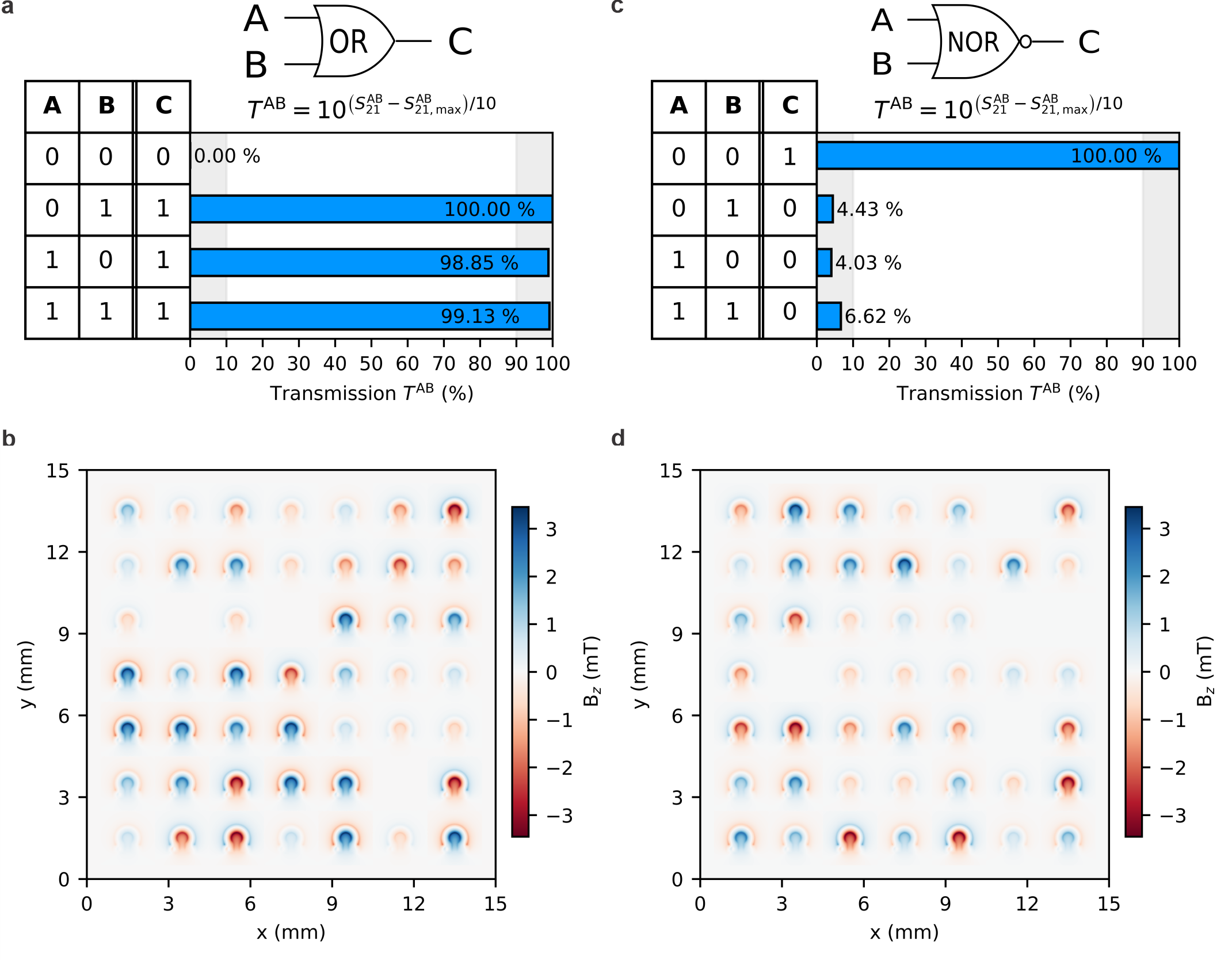}
    \caption{\textbf{OR and NOR logic functionalities}. \textbf{a} \textbf{(OR logic gate)} The spin-wave transmission percentage of output C for all logic combinations of outputs A and B. \textbf{b} A 2D color map representing the Oersted field generated by the omega-shaped loops to achieve results shown in panel a. \textbf{c} \textbf{(NOR logic gate)} The spin-wave transmission percentage of output C for all logic combinations of outputs A and B. \textbf{d} A 2D color map representing the Oersted field generated by the omega-shaped loops to achieve results shown in panel c.}
    \label{fig3}
\end{figure*}

Two inputs and one output are required to perform the OR logic functionality. Input A is assigned to IN1 and input B is assigned to IN3 while output C is assigned to OUT2 -- see Fig.~\ref{fig1}(a). The optimised results of the DS algorithm are presented in Fig~\ref{fig3}(b) applying the following objective function:

\begin{equation}
     O^{\mathrm{OR}} = \left[T^{"01"} + T^{"10"} + T^{"11"}\right]/3,
\end{equation}

where $T^{"01"}$, $T^{"10"}$ and $T^{"11"}$ correspond to the transmission percentages of each input state "AB". The transmission percentages are calculated using Eqn.~\ref{eqn2}. The same conditions apply, where any transmission above 90\% is logic state "1" and transmissions below 10\% are logic state "0". Figure~\ref{fig3}(a) shows the percentage of output transmission for each given input state shown in the OR truth table presented on the left-hand side of the graph. The transmission percentages for the input states "01", "10" and "11" correspond to -46.06\,dB, -46.11\,dB and -46.09\,dB, respectively. In the case of input state "00", no microwave signal is sent to the device and therefore nothing is detected at the output side translating to 0\% transmission or we can consider it as -100\,dB (the smallest transmission reached in~\cite{zenbaa_experimental_2024}), which results in a contrast ratio of at least 53.9. The OR logic gate functionality was achieved in 427 iterations. The field distribution that satisfied the 90/10 transmission condition is presented in Fig.~\ref{fig3}(b).

The NOR gate is known as a universal logic gate because any other logic function (AND, OR, NOT, etc.) can be implemented by combining multiple NOR gates. This makes it possible to construct any digital circuit solely using NOR gates, as they can be cascaded in various configurations to perform any logic operation. It performs the inverse operation of the OR gate, producing a "0" when any of its inputs are "1", in contrast to the OR gate, which outputs "1" when at least one input is "1". It relies on two inputs A and B and one output C and they're again assigned to IN1, IN3 and OUT2 transducers. A feed-line F is needed, like in the case of the NOT gate, and is assigned to IN2. The feed-line F is always open during the measurements. The results of the optimisation are shown in Fig.~\ref{fig3}(c) and were achieved using the following objective function:

\begin{equation}
\begin{split}
     O^{\mathrm{NOR}} = \left[T^{"00"} + (100 - T^{"01"})+ (100 - T^{"10"}) + (100 - T^{"11"})\right]/4.
\end{split}
\end{equation}

The transmission percentages of input states "00", "01", "10" and "11" shown in Fig.~\ref{fig3}(c) correspond to -46.87\,dB, -60.41\,dB, -60.82\,dB and -58.66\,dB, respectively. The contrast ratio $S^{AB}_{21\text{,min trans out"1"}} - S^{AB}_{21\text{,max trans out"0"}}$ is $-46.87 - (-58.66) \approx 11.8$\,dB. The NOR logic functionality was reached after 1461 iterations. Figure~\ref{fig3}(d) shows the Oersted field generated distribution that satisfied the 90/10 condition for a NOR logic functionality.

\subsection{AND and NAND gates}
\begin{figure*}[tbh!]
    \centering
    \includegraphics[width=\linewidth]{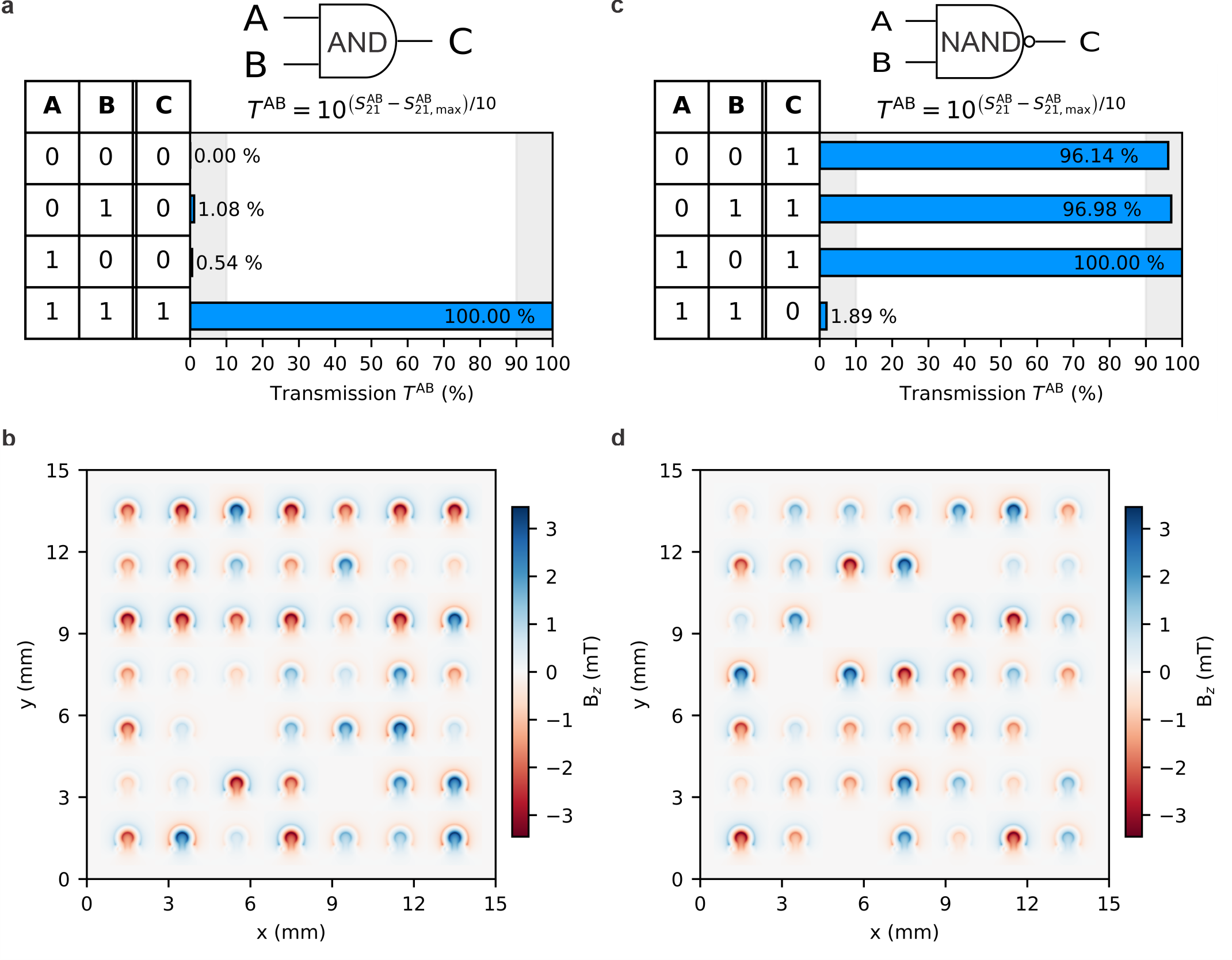}
    \caption{\textbf{AND and NAND logic functionalities}. \textbf{a} \textbf{(AND logic gate)} The spin-wave transmission percentage of output C for all logic combinations of outputs A and B. \textbf{b} A 2D color map representing the Oersted field generated by the omega-shaped loops to achieve results shown in panel a. \textbf{c} \textbf{(NAND logic gate)} The spin-wave transmission percentage of output C for all logic combinations of outputs A and B. \textbf{d} A 2D color map representing the Oersted field generated by the omega-shaped loops to achieve results shown in panel c.}
    \label{fig4}
\end{figure*}

The AND gate is a basic logic gate used in digital circuits. It performs the logical conjunction of its inputs, meaning that it only outputs a logic "1" if all of its inputs are logic "1". If any of the inputs are logic "0", the output will also be logic "0". The AND gate is commonly used in situations where multiple conditions must be true simultaneously for an action to occur, making it an essential component in many computational and control systems. The two inputs A and B are assigned to IN1 and IN3, respectively, and output C to OUT2. The results of the DS optimisation are shown in Fig.~\ref{fig4}(a) using the following objective function:

\begin{equation}
    \begin{split}
     O^{\mathrm{AND}} = \left[(100 - T^{"01"}) + (100 - T^{"10"}) + T^{"11"}\right]/3.
\end{split}
\end{equation}

The transmission percentages of input states "00", "01", "10" and "11" shown in Fig.~\ref{fig4}(a) correspond to -60.99\,dB, -64.007\,dB and -41.33\,dB, respectively. The contrast ratio achieved for the AND gates is $-41.33 - (-60.99) \approx 19.7$\,dB. The AND logic gate was realised in 1118 iterations. The generated Oersted field distribution that achieved the 90/10 conditions is shown in Fig.~\ref{fig4}(b)

The NAND gate is particularly important because it is the second universal gate along NOR, meaning any other logic gate (AND, OR, NOT, etc.) can be constructed using only NAND gates. It performs the inverse operation of the AND gate. Specifically, the NAND gate outputs a logic "0" only when all its inputs are logic "1." For any other combination of inputs, the output will be logic "1." Inputs A and B are assigned to IN1 and IN3, respectively, and output C is assigned to OUT2. Similar to the NOT and NOR logic functionalities, a continuous feed-line F is employed such that it is always put to open and is assigned to IN2. The objective function defined for the NAND gate is as follows:

\begin{equation}
    \begin{split}
        O^{\mathrm{NAND}} = \left[T^{"00"} + T^{"01"} + T^{"10"} + (100 - T^{"11"})\right]/4.
    \end{split}
\end{equation}

The optimised transmission percentages of input states "00", "01", "10" and "11" are 96.14\%, 96.98\%, 100\% and 1.89\%, respectively and are presented in Fig.~\ref{fig4}(c). These percentages correspond to -52.45\,dB, -52.41\,dB, -52.28\,dB and -69.51\,dB. This results in a contrast ratio of at least 17\,dB. The NAND logic gate functionality was reached in 803 iterations. The Oersted field distribution that achieved the 90/10 conditions is shown in Fig.~\ref{fig4}(d).

\subsection{Half-adder logic}

A half adder is a fundamental digital circuit component used to perform the addition of two single binary digits (bits). It has two inputs, typically labeled A and B, and two outputs labeled S (sum) and C (carry). Output S represents the result of the binary addition of the two input bits, while output C accounts for overflow when both input logics are "1". The half adder is typically implemented using an XOR gate for the sum, which outputs "1" if only one of the inputs is "1", and an AND gate for the carry, which outputs "1" when both inputs are "1". In our case, we did not need to implement the half adder using separate XOR and AND gates. Instead, we achieved the truth table output of the half adder directly with our universal device. This approach eliminates the need for a combination of specific logic gates, as our universal device inherently produces the required sum and carry outputs of the half adder truth table. While a half adder can handle the addition of two bits, it does not consider carry inputs from previous operations, making it more suitable for simpler tasks or as a building block for more complex circuits like the full adder, which can manage multi-bit binary addition. Inputs A and B are assigned to IN1 and IN3, respectively, while outputs C and S are assigned to OUT1 and OUT3, respectively. The objective function defined to be maximised by the DS optimiser is as follows:
\begin{equation}
    \begin{split}
     O^{\mathrm{half-adder}} = \left[(100 - T_{\text{C}}^{"01"}) + T_{\text{S}}^{"01"}+(100 - T_{\text{C}}^{"10"})+ \right. \\
     \left.T_{\text{S}}^{"10"} + T_{\text{C}}^{"11"}+(100 - T_{\text{S}}^{"11"})
     \right]/6.
    \end{split}
\end{equation}

\begin{figure*}[tbp!]
    \centering
    \includegraphics[width=\linewidth]{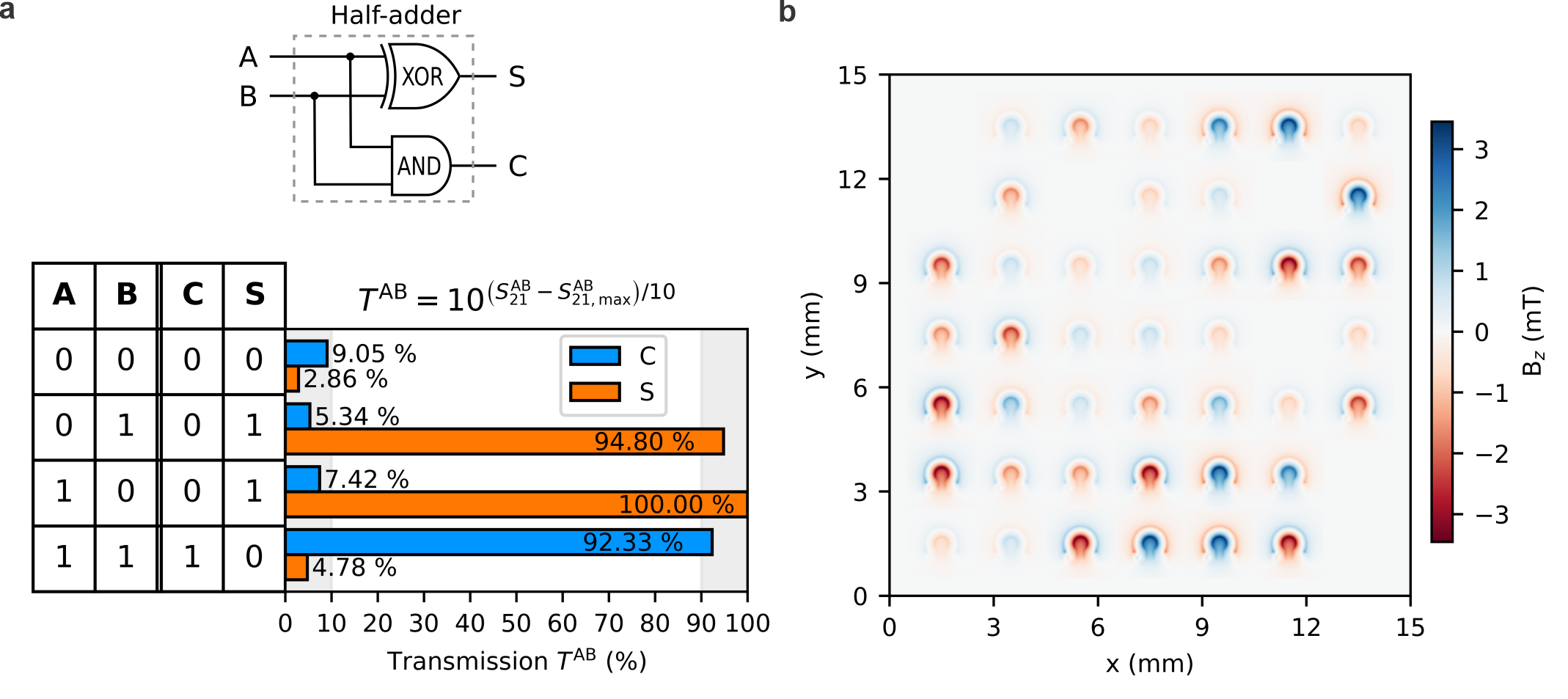}
    \caption{\textbf{The half-adder logic functionality}. \textbf{a} Schematic representation of half-adder gate. \textbf{b} The spin-wave transmission percentage of outputs S and C at all logic combinations of outputs A and B. \textbf{c} A 2D color map representing the Oersted field generated by the omega-shaped loops to achieve results shown in panel b.}
    \label{fig5}
\end{figure*}

where the subscript (C) and (S) correspond to the two outputs C and S, respectively. The transmission percentages for both outputs are given in Fig.~\ref{fig5}(a). Input logic states "01", "10" and "11" yielded transmission percentages of 9.73\%, 7.06\% and 100\% at output C, respectively, satisfying both the truth table shown on the left-hand side of the graph and the 90/10 condition. These percentages correspond to a transmission of -60.84\,dB, -62.24\,dB and -50.73\,dB. Output S transmission percentages are 92.33\% at input logic "01" which corresponds to -51.07\,dB, 96.96\% at input logic "10" which corresponds to -50.86\,dB and 0.98\% at input logic "11" which corresponds to -70.81\,dB. This results in a contrast ratio of about 9.8\,dB. The half-adder logic functionality needed 653 iterations to be realised. The color map of the field distribution generated by the omega-shaped loop array to realise the inverse-design half-adder logic functionality is shown in Fig.~\ref{fig5}(b).

\section{Methods}\label{sec3}
\subsection{Yttrium Iron Garnet sample}
The universal device is based on an 18-µm-thick Yttrium Iron Garnet (YIG) rectangular film with dimensions of 24\,mm$\times$17.5\,mm, grown using Liquid Phase Epitaxy (LPE) on a 500-$\upmu$m-thick Gadolinium Gallium Garnet (GGG) substrate~\cite{dubs_sub-micrometer_2017}. LPE is a growth technique that ensures the YIG film and GGG substrate are lattice-matched, reducing spin-wave damping by promoting optimal crystal alignment.

\subsection{Experimental setup}
\subsubsection{Design region}
The (15$\times$15)\,mm$^2$ design region is comprised of 49 omega-shaped loops and is printed on a printed circuit board (PCB). The width of one loop is about 1.1\,mm with 2\,mm distance between adjacent loops. The PCB consists of four metallic layers, with the design region placed on the top layer, placed on the YIG film. To avoid spin-wave scattering caused by the metallic loops, a Teflon layer was used between the YIG and the PCB. It also served as a heat insulator to prevent heat transfer from the loops to the YIG sample.

\subsubsection{Independent control of current loops}
Five multi-channel current sources, developed by Elbatech Srl., with ten channels each, were used. They were controlled via a PC and designed with feedback loops, enabling precise independent current application in the range of $\pm$1\,A in 2048 steps for each channel.

\subsubsection{Microwave transducers}
Copper microstrip transmission lines, fabricated on a duroid substrate, served as the microwave transducers. These lines are designed with 50\,$\Omega$ impedance at their base and taper down to a width of 50\,$\upmu$m, allowing the excitation of spin waves over a broad wavenumber range, from 3.55\,rad/cm to 0.111\,rad/$\upmu$m~\cite{serga_yig_2010}. The distance traveled by the waves from input to output transducers is about 2.2\,cm.

\subsubsection{Bias magnetic field}
An external magnetic field was applied perpendicular to the sample plane using an electromagnet, with the field strength held at 350\,mT. The field was regulated by a custom-built magnet driver, which continuously monitored and adjusted the field during the experiment, ensuring precision down to $\pm$0.012\,mT. Water-cooled magnet poles were used to dissipate excess heat generated by the PCB carrying the current loops.

\subsubsection{Vector Network Analyzer (VNA)}
To excite and detect spin waves, the setup employed a two-port VNA in conjunction with multiple mechanical switches, allowing signals to be monitored on both the input and output sides of the device. The VNA supported frequency measurements from 10 MHz to 20 GHz.

All data collection and experimental algorithms were implemented using Python libraries.

\subsection{Direct Search (DS) algorithm}
The Direct Search (DS) algorithm, conceptually similar to the Direct Binary Search (DBS) algorithms used in~\cite{wang_inverse-design_2021,shen_integrated-nanophotonics_2015}, differs in that it utilizes a predefined set of finite values, rather than a binary scheme, to identify the optimal solution. The process starts by generating an initial random configuration of currents, denoted as $I_0(n)$, where $n$ represents the number of current loop. This initial configuration consists of 49 current loops, each assigned a random value from the set $S = (i_1, i_2, ...., i_k)$. The algorithm then evaluates the objective function $O$ for the initial state and proceeds by selecting a random loop $n$, checking its current value (e.g., $i_j$), and randomly selecting a different value from the set $S-\{i_j\}$. After updating the current, it recalculates the objective function and compares it to the original $O$. Depending on which value yields a higher objective, it either retains the previous current or applies the new one. This process continues sequentially through all 49 loops, modifying each loop one by one. Once all loops have been altered in this manner, the first iteration is complete, resulting in the updated configuration $I_1(n)$.

\section{Conclusion}\label{sec5}
In this work, we presented the inverse design of all-magnonic logic gates, including NOT, OR, NOR, AND, NAND, and half-adder logic functionalities. They were all realised at a spin-wave frequency of 5.04\,GHz and microwave power of 25\,dBm. By introducing a feed-line F, we made it possible to achieve a logic state "1" at the output for operations like "0 NOT = 1" and "0 NOR 0 = 1". The optimisation procedure used is the DS algorithm. The contrast ratios achieved are 34, 53.9, 11.8, 19.7, 17, and 9.8\,dB for the NOT, OR, NOR, AND, NAND, and half-adder, respectively.

\section*{Acknowledgements}

This research was funded in whole or in part by the Austrian Science Fund (FWF) IMECS [10.55776/PAT3864023]. C.A. acknowledges the support from the Austrian Science Fund (FWF) [10.55776/I6068]. Q.W. acknowledges the support from the National Key Research and Development Program of China (Grant No. 2023YFA1406600). We acknowledge the efforts of ElbaTech Srl in the development of the custom-made multichannel current sources. The authors would like to thank Barbora and Sabri Koraltan for their wedding, resulting in valuable discussions.

\bibliography{ref}

\bibliographystyle{Science}

\clearpage
\renewcommand{\thefigure}{S\arabic{figure}}
\setcounter{figure}{0}
\begin{onecolumn}
\raggedbottom
\begin{center}
    \vfill
    {\LARGE \textbf{Supplementary Information}}
    \vfill
\end{center}
\section*{\underline{\Large Spin-wave full transmission spectrum at 350\,mT}}
Figure~\ref{figS1}a and \ref{figS1}b show the full signal of the spin-wave transmission at 350\,mT. They show the reference signal (at zero applied currents in the omega-shaped loops) of all the input-output combinations of the most complex functionality of a half-adder. Inputs A and B are assigned to IN1 and IN3, respectively, while outputs C and S are assigned to OUT1 and OUT3, respectively. The four spectra show insertion losses that range from -30\,dB to -40\,dB at the working frequency of 5.04\,GHz.

\begin{figure}[hbtp!]
    \centering
    \includegraphics[width=\linewidth]{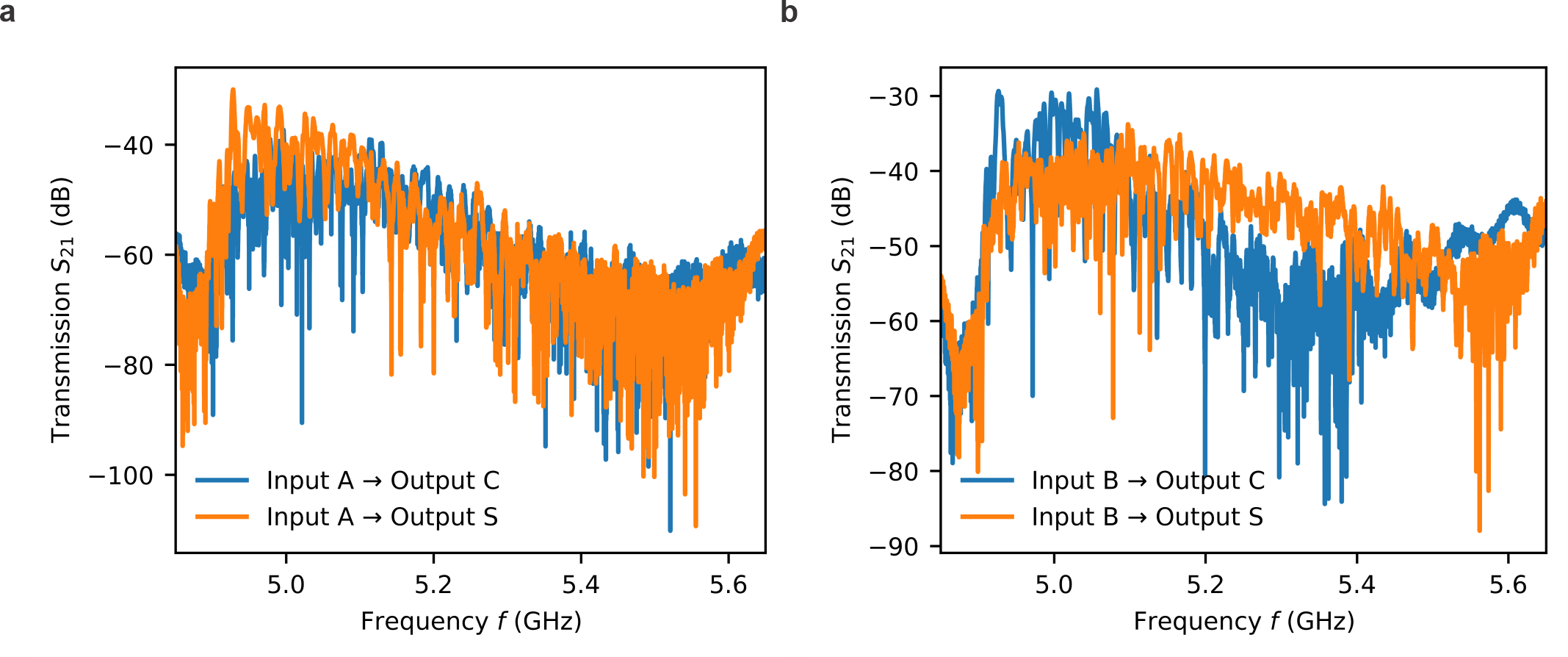}
    \caption{\textbf{Transmission spectra of half-adder at zero-current applied}. \textbf{a} Two transmission spectra from input A to both outputs C and S. \textbf{b} Two transmission spectra from input B to both outputs C and S.}
    \label{figS1}
\end{figure}
\vspace*{\fill}
\end{onecolumn}
\end{document}